\documentclass[12pt]{article} 
\usepackage{graphics} 
\usepackage{latexsym} 
\newcommand{\be}{\begin{equation}} 
\newcommand{\en}{\end{equation}} 
\newcommand{\ee}{\end{equation}}

\newcommand{\ra}{\rightarrow} 
\newcommand{\beq}{\begin{eqnarray}} 
\newcommand{\enq}{\end{eqnarray}} 
\newcommand{\cN}{{\cal{N}}} 
\newcommand{\p}{\partial} 
\newcommand{\cF}{{\cal{F}}} 
\newcommand{\g}{\gamma}
\begin{document} 
\bigskip \bigskip \bigskip 
\centerline{ \bf  \large \large Exotic polarizations of D2 branes and oblique vacua of (S)YM$_{2+1}$} 
\bigskip 

\bigskip 
\centerline{\bf Iosif Bena and Aleksey Nudelman } 
\medskip 
\centerline{Department of Physics} 
\centerline{University of California} 
\centerline{Santa Barbara, CA  93106-9530 U.S.A.} 
\medskip 
\centerline{email:iosif and anudel@physics.ucsb.edu } 
\bigskip \bigskip 
\today
\begin{abstract} 
We investigate the oblique vacua in the perturbed 2+1 dimensional gauge theory living on D2 branes. The 
string theory dual of these vacua is expected to correspond to polarizations of the D2 branes into 
NS5 branes with D4 brane charge. We perturb the gauge theory by adding fermions masses. In the 
nonsupersymmetric case, we also consider the effect of slight variations
of the masses of the scalars. For certain ranges of scalar masses we find oblique vacua.

We show that D4 charge is an essential ingredient in understanding D2 $\ra$ NS5 polarizations. We find that some of 
the polarization states which appear as metastable vacua when D4 charge is not considered are in fact unstable. They
decay by acquiring D4 charge, tilting and shrinking to zero size.
\end{abstract}

\section{\large Introduction}

In the context of the AdS/CFT correspondence \cite{juang}, 
Polchinski and Strassler \cite{joe} used   brane polarization  \cite{myers}, to  find  a string theory dual to a 
confining  four dimensional gauge theory.
Their work has been recently extended to three dimensional  field theories \cite{iosif,us} and to other gauge groups 
\cite{aron}. It was noted in  \cite{us} that near the UV fixed point there might exist oblique vacua corresponding 
to the D2 branes being polarized into an NS5 brane with D4 charge. This might be expected by analogy with the 
oblique vacua of \cite{joe}. Nevertheless, the meaning of this vacua in IIA is more mysterious. Since there is only one 
kind of 1+1 dimensional object (the F string), an oblique vacuum can not be understood as screening dyons. Moreover, the D4 
charge 
of an NS5 ellipsoid is not quantized; it is a dynamical object. This implies that the oblique vacua either 
form a moduli space, or that there is only one isolated oblique vacuum.

The purpose of this paper is to complete the study of D2 brane polarizations, by investigating the existence of such 
oblique vacua. We also study the effect D4 charge (which as we said is dynamic) may have on a D2-NS5 brane 
system. We will
find that a class of D2 $\ra$ NS5 vacua which look  metastable if one does not consider D4 flux, are in fact unstable.
They decay by acquiring D4 charge, tilting, and shrinking to zero size.

Unlike the polarizations of D branes into higher D branes, which can be understood both as a state where the  D 
branes scalars become noncommutative, and as a lower energy configuration with dielectric charge, the D $\ra$ NS5 
polarizations can only be thought of
in the second way. Thus, we study the polarization of D2 branes into NS5 branes with D4 flux by investigating whether a 
wrapped NS5 brane with a very 
large D2 charge and some D4 charge has a ground state at nonzero radius, in the geometry created by itself.

We set three of the fermion masses equal, and we explore several cases. 
In chapter 4 we set the mass of the  fourth fermion to be far less than that of the first three. Our configuration is almost
supersymmetric, and we find a large set of polarizations of D2 branes into NS5/D4 branes, with energies approaching 
0
as $m_4 \ra 0$. We also give the M-theory interpretation to this moduli space.

 In chapter 5 we explore the case of four equal fermions masses, where no supersymmetry is preserved. The 
scalar mass term 
is composed from an $L=0$ piece and various $L=2$ pieces. In the absence of supersymmetry 
we are free to change the $L=2$ mass terms. Chapter 5.1 deals with the SO(4) invariant case, when all the $L=2$ 
pieces are turned off. We find that the vacuum which appears metastable when D4 charge is ignored, is actually unstable. 
It decays by tilting, acquiring D4 charge and finally collapsing into a configuration of zero size.

We then explore the effect of turning on a particular $L=2$ mode. We find a range where polarizations into pure  
NS5 branes are true vacua (chapter 5.2). We also find a configuration where a metastable vacuum corresponding to an 
NS5/D4 polarization is possible (chapter 5.3). Finally in chapter 6 we investigate the meaning of the oblique vacuum 
found in chapter 5.3.
 
\section{\large   The setup}

We perturb the $\cN=8$ theory living on $N$ D2 branes by adding fermion mass terms. By the generalized AdS/CFT
correspondence \cite{imsy} the original theory is dual to string theory in the background given  
by:  
\beq 
ds^2_{string}&=&Z^{-1/2} \eta_{\mu \nu} dx^{\mu} dx^{\nu}+Z^{1/2} dx^{m} dx^{m}, \nonumber \\ 
e^{\phi}&=&g_s Z^{1/4}, \nonumber \\ 
C^0_3&=&-{1 \over g_s Z} dx^0 \wedge dx^1 \wedge dx^2, \ \ \ \ F_4^0 = d C_3^0, 
\label{metric}
\enq 
where $\mu,\nu = 0,1,2$, $m = 3,...,9$, and $g_s$ is the string coupling. When the D2 branes are coincident, 
\be 
Z={R^5\over r^5}, \ \ \ \ R^5 = 6 \pi^2 N g_s {\alpha'}^{5/2}.
\label{2}
\en 
In a general configuration of parallel D2 branes $Z$ is given by the superposition of the individual $Z$'s.
In a similar fashion to the AdS/CFT case \cite{bdhm-bklt}, a fermion mass term in the boundary theory corresponds 
\cite{us} to a nonnormalizable bulk modes of the transverse NS-NS 3 form field strength and the R-R 4 form field 
strength:
\beq 
H_3=g_s\alpha/r^5 (3 T_3-5V_3) = {Z\over 2} (3 T_3-5V_3) , \nonumber \\ 
F^1_4=\alpha/r^5(4 T_4-5V_4) = {Z\over 2g_s}(4 T_4-5V_4) , 
\label{nonnorm} 
\enq 
where the relation between the fermion masses and  $T$ and $V$ are given in the Appendix.

\section{\large The NS5 brane action}
 
In order to study oblique vacua we have to find the action for NS5 branes with D4 brane flux. The field theory living 
on the NS5 brane has a one-form field strength, $\cF$ which comes from the $X^{11}$ coordinate of the M5 brane. 
This field strength couples with the bulk RR $C_5$ via a Wess-Zumino term of the form 
\be
S_{WZ} = \mu_5 \int{\cF \wedge C_5}.
\label{wz-extra}
\ee
Thus a nonzero $\cF$ corresponds to a nonzero D4 brane charge. The general action for an NS5 brane with 
nontrivial $\cF$ flux was found in \cite{sorokin}. 

We expect on general symmetry grounds that the shape of the polarized NS5 brane is a 3-ellipsoid $\times R^3$.
We first examine the action for an NS5 brane shell probe with large D2 charge ($n$) in a background created 
by an even larger number ($N \gg n$) of D2 branes. As we will find, the potential of 
the probe does not depend on the configuration of the $N$ D2 branes. Therefore, this calculation will give, as in 
\cite{joe}, the full self-interacting potential of one or more NS5/D4/D2 shells.

For an SO(3) invariant perturbation ($m_1=m_2=m_3= m$, $0 \neq m_4 \ll m$) we expect the NS5/D4/D2 
configuration to have SO(3) symmetry. Three of the directions of the NS5 brane are extended along the original 
directions of the D2 branes. The other 3 directions are wrapped on an SO(3) invariant 3-ellipsoid. The most general 
SO(3) invariant ellipsoid has its 3 equal axes in  the 3-7, 4-8, and 5-9 planes respectively, at an angle $\g$. The fourth 
axis is along $x^6$. Thus:
\beq 
x^6&=& \alpha\  r \cos \theta, \nonumber \\ 
x^{3'}&=&r \sin \theta \cos \theta_1, \nonumber \\ 
x^{4'}&=&r  \sin \theta \sin \theta_1 \sin \phi, \nonumber \\ 
x^{5'}&=&r  \sin \theta \sin \theta_1 \cos \phi,  
\label{par}
\enq 
where $x^{3'} = x^3 \cos \g+x^7 \sin \g$, $x^{4'}= x^4 \cos \g+x^8 \sin \g $, and $x^{5'}= x^5 \cos \g+x^9 \sin \g $. 
The free parameters are $\g, \alpha$ and $r$. For no D4 charge, we have found \cite{us} a ground state when $\g$ 
was $0$, 
$\alpha$ was $\sqrt{m/m_4}$, and 
\be
r_0^2 = {2 A g_s m \over 3 \alpha},
\label{r0}
\ee
with $A = 4 \pi n (\alpha')^{3/2}$.
A general SO(3) invariant D4 brane charge can be given to the NS5 ellipsoid by turning on a nontrivial
$\cF_{\theta}(\theta)$. Since there is no nontrivial one-cycle on the 3-ellipsoid, the D4 brane charge is not quantized. 
The 
action contains terms of the form \footnote{The sign difference from equation 54 of
\cite{sorokin} is caused by the different metric signature}
$g_{mn} + e^{2 \Phi} \cF_m \cF_n$. The 4-brane charge contributes to the action through 
\be
G_{\theta \theta} \rightarrow G_{\theta \theta} +  e^{2 \Phi} \cF_{\theta} \cF_{\theta} = Z^{1/2} r^2 [\alpha^2 \sin 
^2 \theta \ + \cos^2 \theta + f(\theta)^2],
\label{gtt}
\ee
where we used the metric (\ref{metric}) and defined $f (\theta) \equiv g_s \cF_{\theta}(\theta) /r$. 
In order to give the NS5 brane a D2 charge $n$ we turn on a 3 form field strength along $S^3$
\be
F^{3} ={A} \sin^2 \theta \sin \theta_1\ d \theta \wedge d \theta_1 \wedge d \phi, 
\ee
where $A$ was defined above.
After gauge fixing the NS5 brane action can be written as a combination of a Born-Infeld (BI), mixed, and Wess-
Zumino actions terms:
\beq
{-S_{BI} \over\mu_5 V}  &=& \int{d \theta\ d \theta_1\ d \phi\ {Z^{-1/2} \over g_s^2} \sqrt{G_{\parallel} 
G_{\perp} + g_s^2 Z^{1/2} G_{\parallel} D_{\perp}^2} }, \nonumber \\
{-S_{mixed} \over\mu_5 V}  &=&   - \int{d \theta\ d \theta_1\ d \phi\  {G_{\parallel}  D_{\perp}^2 \over 2 
\sqrt{G_{\parallel} G_{\perp} + g_s^2 Z^{1/2} G_{\parallel} D_{\perp}^2} }},  \label{actions} \\
{-S_{WZ} \over\mu_5 V}&=& - \int{d \theta\ d \theta_1\ d \phi\   [-  B^6_{012 \theta \theta_1\phi}+C^5_{012 
\theta_1\phi} \cF_{\theta}+ {1\over 2} C_{012}F_{\perp} - {1 \over 2} F_{012}C_{\perp}]} \nonumber
\enq
where $D^3=F^3-C^3$, and $G_{\perp}$ is computed using $G_{\theta \theta}$ from (\ref{gtt}). 
The background $C^5$ corresponding to the perturbation (\ref{nonnorm}) is \cite{us}: 
\be
C^5=-\frac{1}{3 g_s} dx^0 \wedge dx^1 \wedge dx^2 \wedge S_2, 
\ee
where $S_2$ is defined in the Appendix. For the fermion masses considered in this paper
\be
\int_{S^2}S_2=4 \pi\ (3m \  {\rm Im}( zz \bar z)+m_4 \ {\rm Im}(zzz)),
\en 
where $z=r e^{i \g}$. Thus, the second term in the Wess-Zumino action is 
\be
\int{d \theta\ d \theta_1\ d \phi\  C^5_{012 \theta_1\phi} \cF_{\theta}} =
-{4 \pi (3m \sin \g + m_4 \sin 3 \g) \over 3 g_s} \int_0^{\pi} (r\sin \theta)^3 \cF (\theta) d\theta,
\ee
where we integrated the $\theta_1$ and $\phi$ components of $C^5$ over a 2-sphere slice of the ellipsoid, of radius 
'$r \sin{\theta}$'. 
As in the simple NS5 case, the equations of motion for large $D_{\perp}$ in the background (\ref{metric}) create an 
$ F_{012}$ which gives a negligible Wess-Zumino contribution.
Using (\ref{metric}, \ref{nonnorm}, \ref{gtt}) and the fact that the D2 brane charge of the 
ellipsoid dominates, we can expand the actions and compute the effective potential. The leading contributions 
in (\ref{actions}) represent D2-D2 interactions and cancel. The potential 
is
\beq
{-S \over\mu_5 V}\!\! &=&\!\! {3 \pi^2 r^6\over 2 Ag_s^3}{3 \alpha^2+1 \over 4} - {\pi^2 \alpha r^4 \over 2 
g_s^2} (3 m 
\cos \g + m_4 \cos (3 \g)) + {\pi^2 A r^2 \over 6 g_s}(3 m^2 + \alpha^2 m_4^2) \nonumber\\
&+&\!\! {3 \pi r^6\over A g_s^3 }   \int_0^{\pi}\sin^2 \theta f(\theta)^2 d\theta - {4 \pi  r^4 (3m\sin \g + m_4 \sin 3 
\g) 
\over 3 g_s^2} \int_0^{\pi}\sin^3 \theta f(\theta) d \theta, \nonumber \\
& &
\label{totalact}
\enq
where the first line is the action with no D4 brane charge, and the second line contains the extra terms appearing 
because of the 4-brane flux. We denote the second part of the action by $S_{D4}$. The last term in the first line 
was obtained by supersymmetric completion of the action with no D4 charge. One might worry that this term changes 
in the presence of D4 charge, and that we should get another term which is the supersymmetric completion of the 
action which includes $S_{D4}$. Nevertheless, we can see from $S_{D4}$ that this new term does not contain 
$f(\theta)$ and reduces to zero when $f=0$, and thus it is zero.

Our unknowns in this potential are $f(\theta),\gamma,\alpha$ and $r$. Our strategy is to first find $f(\theta)$ which 
minimizes $-S_{D4}$, for given $\g,\alpha$, and $r$. We then minimize the resulting function of three variables to 
find the ground state of the NS5/D4/D2 system.

The action $-S_{D4}$ is minimized for 
\be
f(\theta)={4 A g_s (3 m \sin \g + m_4 \sin 3\g)\over 9 r^2} \sin\theta, 
\label{f}
\ee
which gives
\be
{-S_{D4} \over\mu_5 V} = - {\pi^2 m^2 r^2 A  \over 2 g_s} \left({\sin \g + {m_4 \over 3 m} \sin 3\g}\right)^2.
\label{sd4}
\ee

\section{\large The almost supersymmetric case}

In the limit $\chi \equiv m_4/ m \rightarrow 0$, the potential (\ref{totalact}) has a dominant part and a subleading 
part. By minimizing the dominant part we find a relation between $r,\g$ and $\alpha$. We then use this 
relation as a constraint in minimizing the subleading part of potential which disappears as $\chi \rightarrow 0$

For $\chi \ll 1$ we can ignore the terms in (\ref{sd4}) and (\ref{totalact}) which go like $\sin 3 \g $ or $\cos 
3 \g$. Also, in this limit we expect (from the D2/NS5 case) $\alpha$ to be very large, of order $\chi^{-1/2}$.
The dominant part of the potential is:
\be
{-S_{\rm dominant}  \over\mu_5 V} =   {3 \pi^2 r^6\over 2 Ag_s^3}{3 \alpha^2\over 4}
 -  {\pi^2 \alpha r^4 \over 4} (3 m \cos \g)+ {\pi^2 m^2 r^2 A  \over 2 g_s} \cos^2 \g.
\ee
 As expected from supersymmetry, this is a perfect square, and has a minimum for 
\be
\cos \g = {3 \alpha r^2 \over 2 m A g_s}.
\label{gamma}
\ee
In the case when $\g=0$ this formula gives the radius obtained when no $D4$ charge is present. This is consistent 
with (\ref{f}).
The length scale in our problem is set by $m A g_s$, and thus for simplicity we define:
\be
l^2 \equiv  m A g_s
\ee
The subleading part of the potential is:
\be
{-S_{\rm subleading} \over\mu_5 V} = {3 \pi^2\over 8 A g_s^3}
\left[r^6 - {4 \over 3} r^4 l^2 \alpha \chi \cos (3 \g) + {4\over 9}l^4 r^2 (\alpha^2 \chi^2 - 2 \chi \sin \g \sin 
3\g)\right].
\label{sub}
\ee
Combining (\ref{gamma}) with (\ref{sub}) we obtain a potential which gets minimized for $\g=0$, which means no 
D4 charge. The
radius and $\alpha$ are given by (\ref{r0}). 

As $\chi \rightarrow 0$ the potential which fixes $\g$ to zero gets weaker and weaker, and a moduli space opens up. Thus 
as $m_4 \rightarrow 0$, for any radius $r$ smaller than $r_0$, we can 
find $\alpha,\gamma$, and $f(\theta)$ which give zero energy.

This is an interesting result, we find the theory to have a  continuous set of oblique vacua, each corresponding to 
polarization into an NS5 brane with D4 brane charge  given by $f(\theta)$, aspect ratio $\alpha$, radius $r$, and 
orientation $\g$ in the 
3-7, 4-8, and 5-9 planes. When $m_4$ is strictly zero, the long ellipsoids degenerate into lines, so we cannot speak 
properly about NS5 or D4 charges. Nevertheless, this terminology is helpful in describing the D2 branes 
configurations.

There is a very simple  M theory interpretation for these configurations. As can be seen from equation (45) of 
\cite{iosif}, if the fermion mass $m_4$ (linked by SUSY to the coordinates $x^9$ and $x^{10}$) goes to zero, the 
M5 becomes very long
in the $x^9x^{10}$ plane. Moreover, in this limit, the term which depends on the orientation of the M5 brane:
\be
 {\rm Re}(3m\ z_4 zz\bar z+m_4\ \bar z_4 z^3)
\ee
is invariant under $z \ra z e^{i\g}, z_4 \ra z_4 e^{-i\g}$. Thus, we recover a moduli space corresponding to 
this rotation. When reducing these configurations to type IIA, the tilt in the $x^{10}$ direction becomes the D4 flux. 
Therefore we expect to obtain NS5/D4 configurations at an angle $\g$ in the 3-7,5-8 and 6-9 planes, with $D4$ charge
proportional to $\sin \g$, which is exactly what we have in (\ref{f}).
\section{\large Four equal fermion masses}
In this chapter we explore the nonsupersymmetric case when all the fermion masses are equal.
In \cite{joe,us} it was explained that a general scalar mass term, which gives the third term in (\ref{totalact}), is a 
combination of an $L=0$ and an $L=2$ (traceless symmetric) mode:
\beq
 { A\over 12 g_s} (3 m^2) r^2 &=& { A\over 12 g_s} (3 m^2){8 \over 3 \pi}\int_0^{\pi}d \theta \sin^2 \theta  
(x_3^2+x_4^2+x_5^2+ x_7^2+x_8^2+x_9^2) \nonumber\\
&=& { A\over 12 g_s} (3 m^2) {8 \over 3 \pi}\int_0^{\pi} d \theta \sin^2 \theta   \left[{6\over 7} 
(x_3^2+x_4^2+x_5^2+ x_7^2+x_8^2+x_9^2 + x_6^2) \right.  \nonumber \\
&+& \left.  {1\over 7} (x_3^2+x_4^2+x_5^2+ x_7^2+x_8^2+x_9^2-6 x_6^2) \right].
\label{r2}
\enq
The coefficient $(3m^2)$ in front of the $L=0$ mode comes from the square of the fermion masses. When 
$m_4=m$ this is modified to $(4m^2)$.
Supersymmetry fixed the value of the coefficient of the $L=2$ mode in (\ref{r2}), and fixed the coefficients of all other 
$L=2$ components to 0.  In the absence of supersymmetry one can consider adding arbitrary $L=2$ terms.
Since the $L=2$ component in (\ref{r2}) spoils SO(4) invariance everywhere, and moreover does not give any 
nontrivial 
dependence on $\g$ (which is what might give us ground state with nonzero D4 brane charge)  we set it to 0.
We will consider turning on a different  $L=2$ mode of the form:
\be
\Delta S_{L=2} \sim x_3^2+x_4^2+x_5^2+x_6^2 - {3\over4}(x_7^2+x_8^2+x_9^2).
\label{l=2}
\ee
When this mode is absent, all configurations given by (\ref{par}) are SO(4) invariant. When the mode is present only 
configurations in the 3456 plane (that is $\g$=0) are SO(4) invariant. 
The potential is
\be
\frac{-S_{\rm general}}{\mu_5 V}\ \frac{Ag_s^2}{\pi^2} = \frac{3}{2} {3 \alpha^2 +1 \over 4}r^6 - 2 \alpha l^2 
r^4 \cos^3 \gamma +
\label{general}
\ee
\vspace{-.5cm}
\beq 
+\ l^4 r^2 \left[\frac{4}{7} \left(1+{\alpha^2 \over 3}\right)-\frac{1}{2} \left(\sin \gamma +\frac{1}{3} \sin 3 
\gamma \right)^2   + \lambda (4 \sin \g - 3 \cos \g - \alpha^2)\right], \nonumber
\enq
where $\lambda$ is the coefficient of the $L=2$ mode in (\ref{l=2}).

\subsection{The SO(4) invariant case - no polarization}

When the 4 fermion masses are equal and $\lambda=0$, all possible test D2 $\ra$ NS5 configurations 
have SO(4) invariance.
The potential is
\be
\frac{-S}{\mu_5 V}\ \frac{Ag_s^2}{\pi^2} = \frac{3}{2} r^6 - 2 l^2 r^4 \cos^3 \gamma + l^4 r^2 
\left[\frac{16}{21}
-\frac{1}{2} \left(\sin \gamma +\frac{1}{3} \sin 3 \gamma \right)^2 \right].
\label{so4}
\ee
In \cite{us}, we did not take D4 charge into consideration and thus found a local minimum of this potential for 
\beq
\gamma=0 && r=\frac{2\sqrt{g_sA m}}{3} \sqrt{1+ \frac{1}{7}}.
\enq
This minimum has positive second derivative only in the $r$ direction. It is unstable to sliding off in the 
$\gamma$ plane. Thus a test NS5 brane with D2 charge placed in this configuration acquires D4 charge, bends, and
shrinks to zero size.
\begin{center}
\begin{figure}[h]
\scalebox{1.1}{\includegraphics{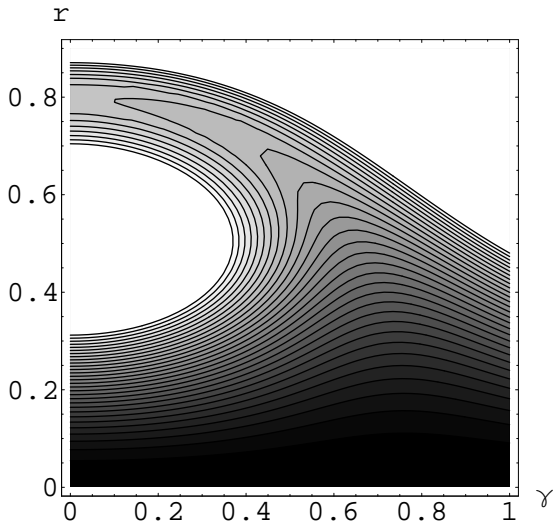}}
\caption{In the SO(4) invariant case, a test NS5 brane lowers its energy by acquiring D4 charge and shrinking to
zero size}
\end{figure}
\end{center}
\vspace{-1cm}
We conclude that the SO(4) invariant theory has no vacuum corresponding to D2 branes polarized into NS5 branes.

\subsection{Polarizations into NS5 branes with no D4 charge}

Turning on an $L=2$ mode of the form (\ref{l=2}) lowers the coefficient of the last term in (\ref{so4}). For the
critical value $\lambda=1/42$, the energy of the $\g=0$ configuration is zero. 
The radius of this 3-sphere ($\g=0$, so as we discussed we have SO(4) invariance) is
\be
r_0={2 \over 3}l.
\ee
Thus, this configuration will not 
shrink to zero size. One might worry that there might be a configuration at some other $\g$ with a lower energy than 
this. 
We plot the potential, and find no other minimum. 
\begin{center}
\begin{figure}[h]
\scalebox{0.9}{\includegraphics{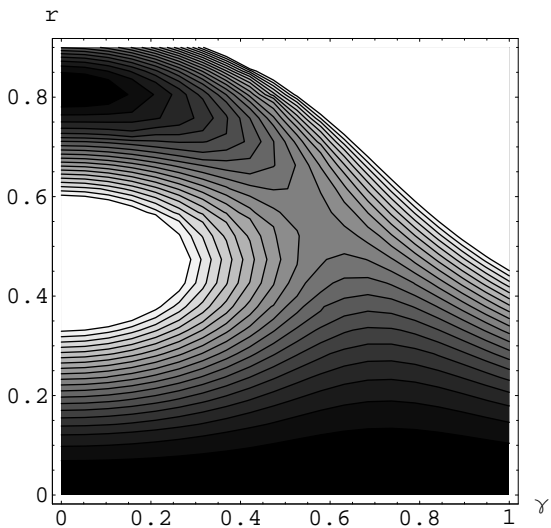}}
\caption{The energy is minimized by a spherical NS5 brane with no D4 charge}
\end{figure}
\end{center}
\vspace{-1cm}
One can also show that the second derivatives of the potential with respect to 
$\g$ and $\alpha$  near the $\g = 0$ zero energy minimum are positive. 
For $\lambda>1/42 $, the energy of the 3-sphere at $\g=0$ is less than 0, so the theory has SO(4) invariant vacua 
corresponding to D2 branes polarized into NS5 branes.

\subsection{Hunting for a minimum with nonzero D4 charge}

For $\lambda = 0$ the lowest energy configuration has zero radius. For $\lambda > 1/42$ the lowest energy 
configuration has nonzero radius and $\g=0$. One might ask the question: Could we have for some 
intermediate $\lambda$ a local minimum with nonzero D4 charge ($\g \neq 0$) ?
To find the minimum of $-S_{rm general}$ we plot it on the hypersurface given by 
\be
{\p S_{\rm general} \over \p \alpha} = 0
\label{hyp}
\ee
as a function of $r$ and $\g$. This hypersurface is given by 
\be
\alpha ={168  r^2 \cos^3 \g \over 32 - 168 \lambda + 189 r^4}
\label{alpha}
\ee
Note that this equation defines a smooth function of $r$ and $\g$ for $\lambda < 8/42$. Since we are exploring the 
range 
$0 < \lambda < 1/42$, the hypersurface is given indeed by a smooth function.
Combining (\ref{alpha}) with (\ref{general}) we obtain a potential which depends on $r$ and $\g$. 
\begin{center}
\begin{figure}[h]
\scalebox{1.1}{\includegraphics{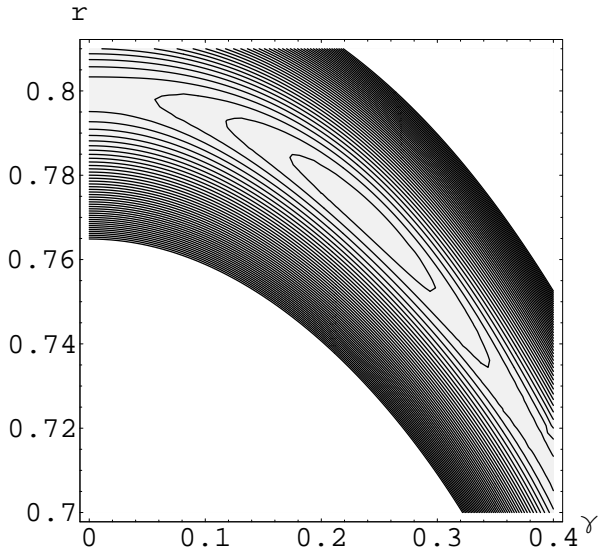}}
\caption{A typical oblique state appears as a local minimum for some values of scalar masses and D4 charge. }
\end{figure}
\end{center}
\vspace{-1cm}
As we lower $\lambda$ from $1/42$, the minimum at $\g=0$ gets higher energy and at about 
$\lambda=.475/42$ starts moving out to nonzero $\g$. For a small interval of $\lambda$'s there is a local 
minimum of energy bigger than zero,  corresponding to
polarization into an NS5 brane with nonzero D4 brane charge. As $\lambda$ decreases further this local minimum 
moves to higher 
$\g$, joins with the true vacuum and disappears.

Thus we have shown that oblique vacua may exist, for some particular values of the $L=2$ scalar masses.
In order to understand the confining properties of this vacuum we need to study its near shell metric.
All the other vacua we found (corresponding to polarizations into pure NS5 branes) are confining (as discussed in 
\cite{us}).

\section{\large The near shell metric}
To obtain a near shell metric for the oblique geometry we start with the NS5 IIB metric
\beq
ds^2_{string} &=& \frac{2 {r_0}_c(\rho^2+\rho^2_c)^{1/2}}{R^2_c} \eta_{\mu \nu} dx^{\mu } dx^\nu+  
\frac{R^2_c}{2 {r_0}_c(\rho^2+\rho^2_c)^{1/2}} (dw^idw^i)  \nonumber\\
&+&\frac{R^2_c 
(\rho^2+\rho^2_c)^{1/2}} {2 {r_0}_c} (dw^3 dw^3 +dy dy), \label{ns5near}  \\
e^{2 \phi} &=& g_s^2 \alpha'^2 \frac{ \sqrt{\rho^2+\rho_c^2} }{\rho^2}, \hspace{1cm} { r_0}_c=m_c \pi 
\alpha' g_s N ,  \nonumber \\
\rho_c &=& \frac{2{r_0}_c \alpha'}{R^2_c}, \hspace{2cm} R^4_c=4 \pi g_s N 
\alpha'^2, \nonumber
\enq
where the $i$'s run from 1 to 2 and $\mu$ and $\nu$ are 0,1,2,3.
Performing a rotation in the $x_3$-$x_4$ plane
\beq
x_3=\cos \gamma x_3'-\sin \gamma  w_1' \nonumber \\
w_1=\sin \gamma x_3'+\cos \gamma w_1'
\enq
we arrive at the metric for a ``tilted''  D3 brane.
We then apply T duality rules of Bergshoeff et al \cite{berg} to get:
\beq
ds^2_{string} &=& h\eta_{\mu \nu} dx^{\mu } dx^\nu+ \frac{ dw_2^2}{h}+\frac{1}{h \cos^2 \gamma +\sin^2 
\gamma/h} (dx_3^{'2}+d w_1^{'2})\nonumber\\
&+&\sin 2 \gamma dx_3^{'} dw_1^{'}(1/h-h)
\frac{R^2_c (\rho^2+\rho^2_c)^{1/2}} {2 {r_0}_c} (dw^3 dw^3 +dy dy), \label{ns5near2} \nonumber  \\
h&=&\frac{2 {r_0}_c(\rho^2+\rho^2_c)^{1/2}}{R^2_c} 
\label{da}
\enq
where $\mu$ and $\nu$ are 0,1,2. The metric (\ref{da}) allows confinement of the
fundamental  string. We could have also obtained this metric by performing a T-duality transformation of the
near shell metric of the oblique vacua of \cite{joe}. The oblique vacuum there confines fundamental charges, 
and screens dyons. Since we have no other objects than fundamental charges in our theory, these vacua 
have no special gauge theory meaning (except being confining), despite their exotic supergravity realization.

\section{\large Discussion}

We have generalized the techniques for exploring D2 brane polarization, by investigating polarization into NS5 branes
with D4 charge. We have found that rather than being an esoteric ingredient, the D4 charge of the NS5 ellipsoids is 
essential in understanding their dynamics, which in its turn gives information about the existence 
of vacua corresponding to polarized branes.
When the mass of one fermion is much smaller than the masses of the other three, we find a moduli space of vacua
corresponding to the D2 branes being polarized into NS5 branes with D4 charge.

When all the fermion masses are equal and all the $L=2$ scalar mass terms are turned off, we have found that the state 
which looks metastable when the D4 charge of the NS5 brane is not considered, is actually unstable. It decays 
by acquiring D4 charge, tilting, and shrinking to zero size. Thus, there are no vacua corresponding to polarized
D2 branes in this case.

Turning on an $L=2$ scalar mass terms allowed us to consider other situations. We have found that for a large enough $L=2$
term a state corresponding to D2/NS5 polarization is energetically stable. Moreover, we found an intermediate
range of strength of the $L=2$ mode where a metastable oblique vacuum appears. We investigated the gauge theory 
properties of this vacuum and found that despite its exotic string theory realization it does not have any
fundamentally different  gauge theory properties.

{\bf Acknowledgements.} 
We are very grateful to Joe Polchinski, Alex Buchel and Amanda Peet for numerous discussions.
The work of I.B. was supported in part by NSF grant PHY97-22022 and the work of A.N. was supported 
in part by DOE contract DE-FG-03-91ER40618.

\section{\large Appendix - Fermion masses and tensor spherical harmonics} 

If we perturb the Lagrangian with the traceless symmetric (in $\lambda_i$) combination 
\be
\Delta L = {\rm Re} (m_1 \Lambda_1^2+ m_2 \Lambda_2^2+ m_3 \Lambda_3^2+ m_4 \Lambda_4^2)
\ee
the corresponding bulk spherical harmonics transforming in the same way under SO(7) are
\beq
T_4& = & {\rm Re} \left(m_1 d\bar z^1 \wedge dz^2 \wedge dz^3 \wedge dx^6+ 
m_2 d z^1 \wedge d\bar z^2 \wedge d z^3 \wedge dx^6 \right. \nonumber \\ 
 &+& \left.m_3 dz^1 \wedge d  z^2 \wedge d\bar z^3 \wedge dx^6+ 
m_4d z^1 \wedge dz^2 \wedge dz^3 \wedge dx^6\right), \nonumber \\ 
T_3&= & {\rm Im} \left(m_1 d\bar z^1 \wedge dz^2 \wedge dz^3 + 
m_2 d z^1 \wedge d\bar z^2 \wedge d z^3  \right. \nonumber \\ 
 &+& \left. m_3 dz^1 \wedge d  z^2 \wedge d\bar z^3 + 
m_4d z^1 \wedge dz^2 \wedge dz^3 \right),
\label{fermass}
\enq
where $z^1=x^3+i x^7,\ \ \ z^2=x^4+i x^8,\ \ \ z^3=x^5+i x^9$.
The bulk spherical harmonics corresponding to fermion masses can be decomposed using the tensors:
\beq 
T_4&=&\frac{1}{4!} T_{mnpk} dx^m \wedge dx^n \wedge dx^p \wedge dx^k \ ,\nonumber \\ 
V_{mnpk}&=&\frac{x^q}{r^2} (x^m T_{qnpk}+x^n T_{mqpk}+x^p T_{mnqk}+x^k T_{mnpq})\ , 
\nonumber \\ 
V_4&=&\frac{1}{4!} V_{mnpk} dx^m \wedge dx^n \wedge dx^p \wedge dx^k\ ,  \nonumber \\ 
S_3&=&\frac{1}{3!}T_{mnpk} x^m \wedge dx^n \wedge dx^p \wedge dx^k\ ,  \nonumber \\ 
T_3&=&\frac{1}{3!}  T_{mnp} dx^m\wedge dx^n\wedge dx^p, \ \ \ \ S_{ 2}=  \frac{1}{2} T_{mnp} 
x^m dx^n\wedge dx^p\ , \nonumber \\
V_{mnp} &=& \frac{x^q}{r^2} ( x^m T_{qnp} + x^n T_{mqp} + x^pT_{mnq}), \ \ \ \  V_{ 3} = 
\frac{1}{3!}  V_{mnp} dx^m\wedge dx^n\wedge dx^p \ , \nonumber \\
\enq

\end{document}